\documentclass[12pt,a4paper]{article}
\usepackage{amsmath} 
\usepackage{mathtools}
\usepackage{amsfonts}
\usepackage{amssymb}
\usepackage{graphicx} 
\usepackage {anysize} 
\usepackage{natbib}
\title{A physically-based model for the electrical conductivity of water-saturated porous media}

\author{Luong Duy Thanh$^{1*}$, {Damien Jougnot}$^2$, Phan Van Do$^1$, Nguyen Van Nghia A$^1$}

\begin{document}
\marginsize{2.5cm}{2cm}{2cm}{2cm}

\date{}
\maketitle

(1) Thuyloi University, 175 Tay Son, Dong Da, Ha Noi, Vietnam
              
(2) Sorbonne Universit\'e,  CNRS, EPHE, UMR 7619 Metis, F-75005, Paris, France 

(*)Corresponding author: thanh{\_}lud@tlu.edu.vn\\

\label{firstpage}

\maketitle

\vskip 10cm

\section*{}
This paper has been published in Geophysical Journal International, please cite as:\\
Luong Duy Thanh, Damien Jougnot, Phan Van Do, Nguyen Van Nghia A, A physically based model for the electrical conductivity of water-saturated porous media, Geophysical Journal International, Volume 219, Issue 2, November 2019, Pages 866--876, https://doi.org/10.1093/gji/ggz328

\newpage

\begin{abstract}

Electrical conductivity is one of the most commonly used geophysical method for reservoir and environmental studies. Its main interest lies in its sensitivity to key properties of storage and transport in porous media. Its quantitative use therefore depends on the efficiency of the petrophysical relationship to link them. In this work, we develop a new physically based model for estimating electrical conductivity of saturated porous media. The model is derived assuming that the porous media is represented by a bundle of tortuous capillary tubes with a fractal pore-size distribution. The model is expressed in terms of the porosity, electrical conductivity of the pore liquid and the microstructural parameters of porous media. It takes into account the interface properties between minerals and pore water by introducing a surface conductivity. Expressions for the formation factor and hydraulic tortuosity are also obtained from the model derivation. The model is then successfully compared with published data and performs better than previous models. The proposed approach also permits to relate the electrical conductivity to other transport properties such as the hydraulic conductivity.
\\

\textbf{Keywords:} Electrical conductivity; Formation factor; Fractal; Porous media
\end{abstract}

\newpage


\section{Introduction}
\label{Introduction}

Measuring electrical conductivity of rocks is of critical importance for oil and mineral exploration \citep[e.g.,][]{PaulGlove2009,Speight} and environmental studies \citep[e.g.,][]{Hubbard2005,RevilKaraoulis2012,doi:10.1002/2015WR017016}. Geoelectrical measurements are a useful, nondestructive tool for characterizing porous rocks and soils \citep[e.g.,][]{BinleyKemna2005,Glover201589}. Conduction of electricity through porous media occurs by two mechanisms: (1) the primary mode of conduction is by movement of ions through the bulk saturating electrolyte and (2) it can also take place in the vicinity of solid surface of pores and cracks \citep[e.g.,][]{doi:10.1190/1.1441549,Revil1997,Revil1999} and that is termed the surface conductance. Surface conductivity includes conduction associated with electrical double layer (EDL): the Stern and diffuse layers and with proton transfer at the interface between the mineral and the pore water \citep[e.g.,][]{Revil1997,Revil1998,Revil1999}. To characterize the relative contribution of the surface conductivity, the dimensionless quantity termed the Dukhin number ($Du$) has been introduced. The Dukhin number is the ratio of the surface conductivity $\sigma_s$ to the pore water conductivity $\sigma_w$ as $Du = {\sigma_s}/{\sigma_w}$ \cite[]{Dukhin1974}. The conductivity of a fully saturated porous medium is related to microstructural properties such as porosity, pore geometry, surface morphology of the mineral grains, pore fluid, temperature \cite[]{FRIEDMAN200545,PaulGlove2009}. For brine-saturated porous materials, \cite{Archie1942} proposed an empirical relationship that links the formation factor $F$ to the porosity $\phi$ based on experimental measurements of electrical conductivity of saturated porous media $\sigma$ and the electrical conductivity of the pore water $\sigma_w$ as follows:

\begin{equation}
\label{eq1}
F=\lim\limits_{\sigma_s\to 0}\left(\frac{\sigma_w}{\sigma}\right)=\phi^{-m},\\
\end{equation}
where $m$ is called the cementation exponent that is supposed to be constant for a certain type of rock. The cementation exponent $m$ is normally between 1.3 and 2.5 for unconsolidated sands, porous sediments and higher than 2.5 for carbonate rocks where the pore space is not well connected \citep[e.g.,][]{FRIEDMAN200545,PaulGlove2009,GLOVER199757}. Archie${'}$s law is playing an important role in the exploration of oil reservoir \citep[e.g.,][]{PaulGlove2009,JCarpenter2009}. The electrical conductivity $\sigma$ of saturated porous media for negligible surface conductivity can be obtained from Eq. (\ref{eq1}) as
\begin{equation}
\label{eq2}
\sigma = \sigma_w\phi^{m}.
\end{equation}

Beside the Archie model, many models are available in literature \citep[e.g.,][]{FRIEDMAN200545,Laloy2011,CAI2017419} to link the electrical conductivity of saturated porous media $\sigma$ and the electrical conductivity of the pore fluid $\sigma_w$. For example, many relationships between $\sigma$ and $\sigma_w$ have been obtained from the effective medium theories \citep[e.g.,][]{doi:10.1190/1.1441549,doi:10.1190/1.1441215,doi:10.1190/1.1442295,doi:10.1190/1.3483875}, the percolation theory \cite[]{GRL:GRL18768,Ghanbarian2014}, the cylindrical tube model \cite[]{doi:10.1190/1.1443651}, among other approaches. Existing electrical models with surface conductivity based on different approaches such as the two-resistors in parallel approach \citep[e.g.,][]{Waxman1968,Brovelli2005}, the effective medium \citep[e.g.,][]{Hanai1960,doi:10.1190/1.1441549}, the volume averaging theorem \citep[e.g.,][]{Pride,doi:10.1029/2006WR005131} have also been presented in literature. Recently, fractal models have been proven to be an alternative and useful means for studying the transport phenomenon and analyzing the macroscopic transport properties of porous media \citep[e.g.,][]{Mandelbrot1982,doi:10.1080/00018738700101062,Feder1989,doi:10.1146/annurev.ea.19.050191.001321,RevModPhys.65.1393,doi:10.4141/cjss10008,PENGXU2015,Luong2018,doi:10.1002/2017JB014873}. Fractal theory in porous media has been applied to derive theoretical electrical conductivity models \citep[e.g.,][]{PhysRevLett.54.1325,PapeH1987,PhysRevB.55.8038,PhysRevE.78.016308,PhysRevLett.100.035504,GRL:GRL53088}. However, to the best of our knowledge, the surface electrical conductivity is not yet taken into account in the fractal electrical conductivity models that are already published. Therefore, one may underestimate electrical conductivities of saturated porous media and fail to reproduce experimental data when the electrical conductivity of the pore fluid is low. 

In this work, we propose a fractal model for the electrical conductivity of saturated porous media based on the fractal theory of porous media and on the capillary bundle model. The proposed model is expressed in terms of microstructural parameters such as the minimum and maximum pore/capillary radii, the pore fractal dimension, the tortuosity fractal dimension and the porosity of porous media. The model prediction is then compared with experimental data in the literature. A good agreement is found between the prediction by the proposed fractal model and experimental data. Factors influencing the electrical conductivity in porous media are also analyzed.
\begin{figure}
  \centering
  \includegraphics[width=7.0cm]{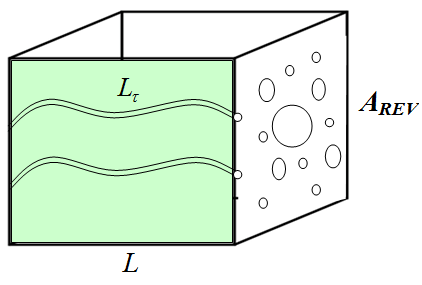}
  \caption{A porous rock model composed of a large number of parallel capillary tubes with radius following a fractal distribution.}\label{fig1}
\end{figure}
\section{Fractal theory for porous media}
\label{Fractal porous theory}

Fractal theory for porous media has been already presented by many works \citep[e.g.,][]{BYu2002,Liang2014,Liang2015} in which a porous medium is assumed to be made up of an array of tortuous capillaries with different sizes (see Fig. \ref{fig1}). The pore size distribution in porous media has been proven to follow the fractal scaling law

\begin{equation}
\label{eq3}
N(\geq r) =(\frac{r_{max}}{r})^{D_f},
\end{equation}
where $N$ is the number of capillaries (whose radius $\geq$ r) in a fractal porous media, $r_{max}$ is the maximum radius of capillary, $D_f$ is the fractal dimension for pore space, 0 $<$ $D_f$ $<$ 2 in two-dimensional space and 0 $<$ $D_f$ $<$ 3 in three dimensional space \cite[]{BYu2002,Liang2014,Liang2015}. Eq. (\ref{eq3}) implies the property of self-similarity of porous media, which means that the value of $D_f$ from Eq. (\ref{eq3}) remains constant across a range of length scales. It should be noted that porous media are not exactly self-similar. Porous media are shown to be statistically self-similar because they exhibit the self-similarity in some average sense and over a certain range of length scales \citep[e.g.,][]{doi:10.1142/S0218348X01000610,BYu2002}. As there are numerous capillaries in porous media, Eq. (\ref{eq3}) can be considered as a continuous function of the radius.\\ 

Differentiating Eq. (\ref{eq3}) with respect to $r$ yields 
\begin{equation}
\label{equa4}
-dN=D_f r_{max}^{D_f} r^{-D_f-1} dr,
\end{equation}
where -$dN$ represents the number of pores from the radius $r$ to the radius $r+dr$.  The minus (-) in Eq. (\ref{equa4}) implied that the number of pores decreases with the increase of pore size. Therefore, the total number of capillaries from the minimum radius $r_{min}$ to the maximum radius $r_{max}$ can be obtained by 

\begin{equation}
\label{equa5}
N_{total}(\geq r_{min}) =(\frac{r_{max}}{r_{min}})^{D_f}.
\end{equation}

From Eq. (\ref{equa4}) and Eq. (\ref{equa5}), we obtain
\begin{equation}
\label{equa6}
-\frac{dN}{N_{total}}=D_f r_{min}^{D_f} r^{-D_f-1} dr = f(r) dr,   
\end{equation}
where $f(r) = D_f r_{min}^{D_f} r^{-D_f-1}$ is the probability density function of the pore size distribution. This function must satisfy the following condition

\begin{equation}
\label{equa7}
\int_{r_{min}} ^{r_{max}} f(r) dr = 1-(\frac{r_{min}}{r_{max}})^{D_f} = 1
\end{equation}
or
\begin{equation}
\label{equa8}
(\frac{r_{min}}{r_{max}})^{D_f} \approx 0.
\end{equation}
As stated in \cite{BYu2002}, \cite{Liang2014} and \cite{Liang2015}, for example, Eq. (\ref{equa8}) holds approximately when $r_{min}/r_{max}$ $\approx$ 10$^{-2}$ or $<$ 10$^{-2}$ in porous media. Generally, the above condition holds in porous media. Therefore, the fractal theory can be used to investigate properties of porous media \citep[e.g.,][]{PhysRevLett.54.1325,doi:10.1142/S0218348X01000610,YU20022983}. The fractal dimension for pore space ($D_f$) is expressed as \citep[e.g.,][]{doi:10.1142/S0218348X01000610,BYu2002} 

\begin{equation}
\label{equa9}
D_f=2-\frac{\mathrm{ln} \phi}{\mathrm{ln} \alpha},
\end{equation}
where $\phi$ is the porosity of porous media and $\alpha$ is the ratio of the minimum pore radius to the maximum pore radius ($\alpha$ = $r_{min}/r_{max}$).\\

\section{Theoretical development of a new electrical model}
\label{electricalmodel}
\subsection{Pore scale}
\label{Porescale}

If the capillary radius is $r$ and its length is $L_{\tau}$ and if it is filled with a fluid of conductivity $\sigma_w$, then the resistance ($R_c$) of the capillary should be \cite[]{Norman1961}

\begin{equation}
\label{equa11}
\frac{1}{R_c}= \frac{\pi r^2\sigma_w}{L_{\tau}},
\end{equation}

However, if there is a contribution from surface conductance (specific surface conductance $\Sigma_s$), then the resistance ($R_s$) due to the surface conductance (i.e. the EDL) would be

\begin{equation}
\label{equa12}
\frac{1}{R_s}= \frac{2\pi r\Sigma_s}{L_{\tau}},
\end{equation}
It should be noted that $\Sigma_s$ is related to the cationic exchange capacity of the porous medium \cite[]{doi:10.1029/98JB02125,doi:10.1029/2011WR010919}.

The total resistance of the capillary (the two conductors in parallel) is given by \cite[see][for more details]{PFANNKUCH197242}

\begin{equation}
\label{equa13}
\frac{1}{R(r)}= \frac{1}{R_c}+\frac{1}{R_s}=\frac{\pi r^2\sigma_w}{L_{\tau}(r)}+\frac{2\pi r\Sigma_s}{L_{\tau}(r)}.
\end{equation}
One of the assumptions behind Eq. (\ref{equa13}) is the percolation of the surface-conductive layer. This assumption is made in widely used models from the literature \citep[e.g.,][]{Waxman1968,Pride,Gueguen1994,doi:10.1029/2006WR005131,doi:10.1029/98JB02125}. 

When this is not the case, one has to use different models such as self-similar up-scaling techniques \citep[e.g.,][]{Sen} or the percolation theory \citep[e.g.,][]{Ghanbarian2014}.

The relationship between the radius and length of the capillary should conform to the following fractal scaling relationship \citep[e.g.,][]{YU20022983,WU20073925}
\begin{equation}
\label{equa14}
L_{\tau}(r)= r^{1-D_{\tau}}L_o^{D_{\tau}},
\end{equation}
where $D_{\tau}$ is the fractal dimension for the  geometrical tortuosity of the capillaries with 1 $<$ $D_{\tau}$ $<$ 2 in two dimensions, $L_o$ is the length of the porous media and $L_{\tau}$ is the tortuous length. 

From Eq. (\ref{equa14}), the geometrical tortuosity of a capillary having a radius $r$ is obtained as
\begin{equation}
\label{equa15}
\tau(r)=\frac{L_{\tau}(r)}{L_o}=(\frac{L_o}{r})^{D_{\tau}-1}.
\end{equation}
Eq. (\ref{equa15}) indicated that smaller diameter capillaries are more tortuous than larger ones. That is consistent with the physical situation for an artery and micro-blood vessel. The artery is much larger in diameter but much less tortuous than the blood vessel \cite[]{0256-307X-22-1-045,Doyen1988}. It should be noted that at the microscale, tortuosities can be either geometrical or electrical, or hydraulic. They are dependent of the capillary size as shown by Eq. (\ref{equa15}). At the microscopic scale, we define the geometrical tortuosity ($\tau_g$) as the ratio between the effective shortest possible length between inflow and outflow points that avoid solid grains (this is realized as a zigzag path passing grains with close tangents) and the straight line length in the direction of flow in porous media \cite[]{Clennell299}. Hydraulic tortuosity ($\tau_h$) is defined as the ratio of the effective fluid path length between solid grains (this is realized as a smoothed route through porous media)and the straight line length in the direction of flow \cite[]{Clennell299}. Electrical tortuosity ($\tau_e$) is defined as the ratio of the effective path length for electrical flow and the straight line length of porous media \cite[]{Clennell299}. It is seen that $\tau_g < \tau_e<\tau_h$ \cite[]{GhanbarianBehzad2013}. Note that, at the macroscale scale (i.e., REV scale described in the next section), these tortuosities become effective tortuosities (denoted $''eff''$): geometrical ($\tau^{eff}_g$), electrical ($\tau^{eff}_e$), and hydraulic ($\tau^{eff}_h$). That is, they result from the analytical up-scaling of the tortuosities of many capillaries sizes described in the next section.

\subsection{REV scale}
\label{REVscale}
In order to derive the electrical conductivity at macroscale, we consider a representative elementary volume (REV) as a cube with 
the length of $L_o$ and the cross-section area of the REV perpendicular to the flow direction of $A_{REV}$. The porous medium of the REV is conceptualized as an equivalent bundle of capillary tubes with a fractal pore size distribution and the pore structure with radii varying from a minimum pore radius $r_{min}$ to a maximum pore radius $r_{max}$ as described in the previous section.

The resistance of the water-saturated rock in the REV can be obtained as \cite[]{Jackson2008,Jackson2010,Wang2014}
\begin{equation}
\label{equa16}
\begin{split}
\frac{1}{R_o} =&\int_{r_{min}} ^{r_{max}} \frac{1}{R(r)} (-dN)\\
		=&\int_{r_{min}} ^{r_{max}} (\frac{\pi r^2\sigma_w}{r^{1-D_{\tau}}L_o^{D_{\tau}}}+\frac{2\pi r\Sigma_s}{r^{1-D_{\tau}}L_o^{D_{\tau}}}) D_f r_{max}^{D_f} r^{-D_f-1} dr\\
		=&\frac{\pi \sigma_w D_f r_{max}^{D_{\tau}+1}}{L_o^{D_{\tau}}.(D_{\tau}-D_f+1)}.(1-\alpha^{D_{\tau}-D_f+1})+\frac{2\pi \Sigma_s D_f r_{max}^{D_{\tau}}}{L_o^{D_{\tau}}.(D_{\tau}-D_f)}.(1-\alpha^{D_{\tau}-D_f})\\
\end{split}
\end{equation}
Additionally, the effective resistance of the saturated porous medium at the REV scale can be calculated as 
\begin{equation}
\label{equa17}
R_o= \frac{L_o}{\sigma A_{REV}} 
\end{equation}
where $\sigma$ is the electrical conductivity of the water-saturated porous media.

The porosity of the REV is defined as the ratio of the total pore volume $V_p$ and the total volume of the REV $V_{REV}$ \citep[e.g.,][]{Jackson2008,Jackson2010}:
\begin{equation}
\label{equa18}
\phi=\frac{V_p}{V_{REV}}\\
\end{equation}
Therefore, the porosity is calculated as
\begin{equation}
\label{equa19}
\begin{split}
\phi=&\frac{\int_{r_{min}} ^{r_{max}} \pi r^2 L_{\tau}(r)(-dN)}{A_{REV}L_o}=\frac{\int_{r_{min}} ^{r_{max}} \pi r^2 r^{1-D_{\tau}}L_o^{D_{\tau}} D_f r_{max}^{D_f} r^{-D_f-1} dr}{A_{REV}L_o}\\
    =&\frac{\pi L_o^{D_{\tau}-1} D_f r_{max}^{3-D_{\tau}}} {A_{REV}(3-D_{\tau}-D_f)}(1-\alpha^{3-D_{\tau}-D_f})\\
\end{split}
\end{equation}
Consequently, the following is obtained
\begin{equation}
\label{equa20}
A_{REV} =\frac{\pi D_f L_o^{D_{\tau}-1} r_{max}^{3-D_{\tau}}}{\phi}\frac{(1-\alpha^{3-D_{\tau}-D_f})}{(3-D_{\tau}-D_f)}\\
\end{equation}
Combining Eq. (\ref{equa16}), Eq. (\ref{equa17}) and Eq. (\ref{equa20}) yields
\begin{equation}
\label{equa21}
\begin{split}
\sigma=&\frac{\phi r_{max}^{2D_{\tau}-2} (3-D_{\tau}-D_f).(1-\alpha^{D_{\tau}-D_f+1})}{{L_o}^{2D_{\tau}-2}.(D_{\tau}-D_f+1).(1-\alpha^{3-D_{\tau}-D_f})}.\sigma_w\\
			+&\frac{2\phi r_{max}^{2D_{\tau}-3} (3-D_{\tau}-D_f).(1-\alpha^{D_{\tau}-D_f})}{{L_o}^{2D_{\tau}-2}.(D_{\tau}-D_f).(1-\alpha^{3-D_{\tau}-D_f})}.\Sigma_s\\
\end{split}
\end{equation}
This equation is the main contribution of this work. Eq. (\ref{equa21}) indicates that the electrical conductivity of porous media under saturated conditions is explicitly related to the porosity, electrical conductivity of the pore liquid, and the microstructural parameters of a porous medium ($D_f$, $D_{\tau}$, $\phi$, $\alpha$, $r_{max}$). Therefore, the model can reveal more mechanisms affecting the electrical conductivity of saturated porous media than other models available in literature. In particular, there is no empirical constant in Eq. (\ref{equa21}) such as constants $F$ or $m$ in Archie model, for example. Eq. (\ref{equa21}) can be rewritten as

\begin{equation}
\label{equa22}
\begin{split}
\sigma=&\frac{\phi (3-D_{\tau}-D_f)(1-\alpha^{D_{\tau}-D_f+1})}{(\tau^{eff}_g)^2.(D_{\tau}-D_f+1).(1-\alpha^{3-D_{\tau}-D_f})}.\sigma_w\\
			+&\frac{2}{r_{max}}\frac{\phi}{(\tau^{eff}_g)^2}\frac{(3-D_{\tau}-D_f)}{(D_{\tau}-D_f)}\frac{(1-\alpha^{D_{\tau}-D_f})}{(1-\alpha^{3-D_{\tau}-D_f})}.\Sigma_s,\\
\end{split}
\end{equation}
where $\tau^{eff}_g$ is the effective geometrical tortuosity of the medium resulting from the contributions of all the capillary tortuosities as presented above and is given by
\begin{equation}
\label{equa23}
\tau^{eff}_g =(\frac{L_o}{r_{max}})^{D_{\tau}-1}.
\end{equation}

Eq. (\ref{equa22}) has the similar form to classical models reported by many authors \citep[e.g.,][]{Waxman1968,Pride,doi:10.1029/98JB02125,FRIEDMAN200545,doi:10.1029/2006WR005131,PaulGlove2009,doi:10.1029/2011WR010487} as following:
\begin{equation}
\label{equa24}
\begin{split}
\sigma=g.\sigma_w+h.\sigma_s\\
\end{split}
\end{equation}
where $\sigma_s$ is the surface conductivity of porous media, $g$ and $h$ are two dimensionless geometrical factors, which depend on the texture of porous media.
 
The length of representative volume unit is related to the cross-section area of the REV by \cite[]{MIAO2016884}
\begin{equation}
\label{equa25}
{L_o}^2 =A_{REV}\\
\end{equation}
Note that $L_o$ is not the length of sample but the side length of a representative unit as the REV is assumed to be a cube.\\ 
From Eq. (\ref{equa23}) and Eq. (\ref{equa25}), one has

\begin{equation}
\label{equa26}
\tau^{eff}_g ={\left[{\frac{1-\alpha^{3-D_{\tau}-D_f}}{\phi}\frac{\pi D_f}{3-D_{\tau}-D_f}}\right]}^{\frac{D_{\tau}-1}{3-D_{\tau}}}\\
\end{equation}
The fractal dimension for the tortuosity $D_{\tau}$ can be approximately expressed as a function of properties of porous media as \cite[]{GRL:GRL53088}
\begin{equation}
\label{equa27}
D_{\tau} =(3-D_f)+(2-D_f)\frac{\mathrm{ln}\frac{D_f}{D_f-1}}{\mathrm{ln}\phi}
\end{equation}

When the surface conductivity is negligible, Eq. (\ref{equa22}) becomes
\begin{equation}
\label{equa30}
\begin{split}
\sigma=&\frac{\phi (3-D_{\tau}-D_f)(1-\alpha^{D_{\tau}-D_f+1})}{(\tau^{eff}_g)^2 (D_{\tau}-D_f+1)(1-\alpha^{3-D_{\tau}-D_f})}.\sigma_w\\
\end{split}
\end{equation}

For a porous medium made up of straight parallel capillaries containing a fluid, $D_{\tau}$ approaches 1. Therefore, the parameter $\tau^{eff}_g$ is equal to 1 as indicated by Eq. (\ref{equa23}) and Eq. (\ref{equa30}) becomes $\sigma = [\phi (2-D_f)(1-\alpha^{2-D_f})\sigma_w]/[(\tau^{eff}_g)^2 (2-D_f)(1-\alpha^{2-D_f})]$=$\phi\sigma_w/(\tau^{eff}_g)^2=\phi.\sigma_w$. This prediction is exactly what one would expect and is identical to the parallel model with one conducting phase and one insulating phase given by \cite{Gueguen1994}. Under that condition, Eq. (\ref{equa30}) also corresponds to Archie$'s$ law by considering $\phi$ = 1.\\
From Eq. (\ref{eq1}) and Eq. (\ref{equa30}), the formation factor is obtained as
\begin{equation}
\label{equa31}
\begin{split}
F=&\frac{(\tau^{eff}_g)^2}{\phi}\frac{(D_{\tau}-D_f+1)(1-\alpha^{3-D_{\tau}-D_f})}{(3-D_{\tau}-D_f)(1-\alpha^{D_{\tau}-D_f+1})}.\\
\end{split}
\end{equation}

Similarly, Eq. (\ref{equa31}) indicates that the formation factor is related to the porosity, and the microstructural parameters of a porous medium ($D_f$, $D_{\tau}$, $\phi$, $\alpha$). Additionally, Eq. (\ref{equa31}) shows that the formation factor is inversely proportional to the porosity in the similar form as reported in \cite{Clennell299} (see his Eq. (17) and reference therein). It is seen from Eq. (\ref{equa31}) that the formation factor can be considered independent of the grain size as indicated in the experimental results of \cite{Glover2010}, as long as the other geometrical parameters remain constant.

Even though the present model is based on the concept of capillary tubes, one can relate capillary radius to grain size. Indeed, in non consolidated granular materials, grain sizes and grain size distribution are much easier to obtain than pore sizes. Therefore, if a granular material is considered, \cite{Cai2012} proposed an expression to calculate maximum radius of the capillaries by combining the geometrical models of an equilateral-triangle and a square arrangement of spherical grains as
\begin{equation}
\label{equa10}
r_{max} = \frac{d}{8} \left[\sqrt{\frac{2\phi}{1-\phi}} + \sqrt{\frac{\phi}{1-\phi}}+\sqrt{\frac{\pi}{4(1-\phi)}}-1\right],
\end{equation}
where $d$ is the mean grain diameter in porous media. Eq. (\ref{equa10}) predicts that (1) when porosity approaches zero, $r_{max}$ turns negative; (2) when porosity is approximately greater than 0.67, $r_{max}$ becomes larger than mean grain radius. Those predictions are unrealistic. However, porosity of porous media is never zero and normally reported to be less than 0.45 \citep[e.g.,][]{Waxman1968}. Therefore, the pore radius is typically less than grain radius as expected from the literature \citep[e.g.,][]{Glover2009}.

\section{Results and discussion}
\label{Results}


\begin{table*}
\caption{Some of the models for the electrical conductivity of porous media used in this work for comparison. It should be noted that $\sigma_m$ is the electrical conductivity of the solid matrix and taken as zero in this work.}
\begin{center}
\begin{tabular}{c c c }
  \hline
  Name&Equation& Reference\\
  \hline
  Archie model & $\sigma= \sigma_w\phi^{m}$ & \cite{Archie1942} \\
	Parallel model & $\sigma= \phi\sigma_w+(1-\phi)\sigma_m$ & \cite{Gueguen1994}  \\
	Waff model & $\sigma= \frac{\sigma_w+(\sigma_m-\sigma_w)(1-2\phi/3)}{1+\frac{\phi}{3}(\frac{\sigma_m}{\sigma_w}-1)}$ & \cite{doi:10.1029/JB079i026p04003} \\
	Pride model & $\sigma= \frac{1}{F}\left[\sigma_w+(F-1)\sigma_s\right]$ & \cite{Pride} \\
	Revil model & $\sigma= \frac{\sigma_w}{F}[1-t^f_{(+)} + F.Du+\frac{1}{2}(t^f_{(+)}-Du)$ & \cite{doi:10.1029/98JB02125}  \\
	 & $\times(1-\frac{Du}{t^f_{(+)}}+\sqrt{(1-\frac{Du}{t^f_{(+)}})^2+4F\frac{Du}{t^f_{(+)}}})]$ &  \\
	\hline
\end{tabular}
\end{center}
\label{table2}
\end{table*}
\subsection{Sensitivity analysis of the new model}
\label{Sensitivitymodel}

For modeling the electrical conductivity of saturated porous rocks as a function of the pore fluid electrical conductivity based on Eq. (\ref{equa22}), one needs to know the fractal parameters $\alpha$, $D_f$, $r_{max}$, $\tau_{min}$ and $D_{\tau}$. It was generally recognized that the minimum pore radius is approximately two orders of magnitude smaller than the maximum pore radius in porous media. The value of $\alpha$ = 0.01 has been used in published articles \citep[e.g.,][]{Cai2012,Liang2014,Liang2015}. Therefore, the value $\alpha$ = 0.01 is also used in this work for modeling. The fractal dimension $D_f$ is determined via Eq. (\ref{equa9}). If the pore size distribution of porous media is not known, the maximum radius $r_{max}$ can be estimated via Eq. (\ref{equa10}). $D_{\tau}$ is obtained from Eq. (\ref{equa27}) with the knowledge of $D_f$ and $\phi$. From Eq. (\ref{equa26}), $\tau^{eff}_g$ is then determined. Substituting $r_{max}$, $D_f$, $D_{\tau}$ and $\tau^{eff}_g$ into Eq. (\ref{equa22}), the electrical conductivity of porous samples is then determined with known values of $\sigma_w$ and $\Sigma_s$. 

\begin{figure}
  \centering
  \includegraphics[width=11.0cm]{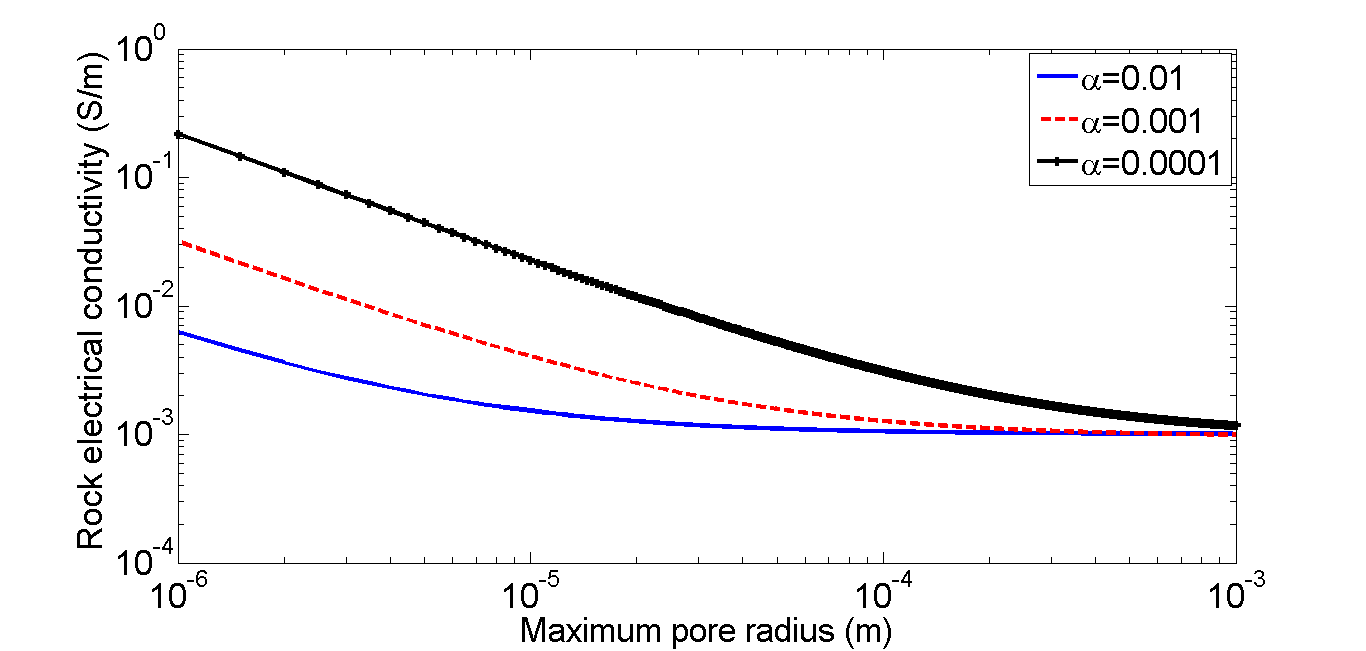}
  \caption{The variation of the electrical conductivity of porous samples with the maximum pore radius predicted from the model presented by Eq. (\ref{equa22}) for three different values of $\alpha$ (0.01, 0.001 and 0.0001) with $\sigma_w$ = 3.0$\times10^{-3}$ S/m, $\Sigma_s$ = 0.5$\times10^{-9}$ S and $\phi$ = 0.4.}\label{Figure2}
\end{figure}

Fig. \ref{Figure2} shows the variation of the electrical conductivity of porous media with the maximum pore radius predicted from the model presented by Eq. (\ref{equa22}) for three different ratios of the minimum radius to the maximum radius of capillaries ($\alpha$ = 0.01, 0.001 and 0.0001) with $\sigma_w$ = 3.0$\times10^{-3}$ S/m, $\Sigma_s$ = 0.5$\times10^{-9}$ S and $\phi$ = 0.4. It can be seen that when the ratio $\alpha$ decreases, the electrical conductivity increases. This is attributed to the fact that the smaller ratio $\alpha$ at given porosity causes the higher fractal dimension for pores as seen in Eq. (\ref{equa9}). Additionally, the electrical conductivities of porous media for three different values of $\alpha$ approach the constant and the same value at large grain diameters at which the surface electrical conductivity is negligible. This observation is in good agreement with what is predicted from Archie${'}$s law $\sigma= \sigma_w\phi^{m}$ in which the electrical conductivity of porous media only depends on the porosity regardless of the grain size.

\begin{figure}
  \centering
  \includegraphics[width=11.0cm]{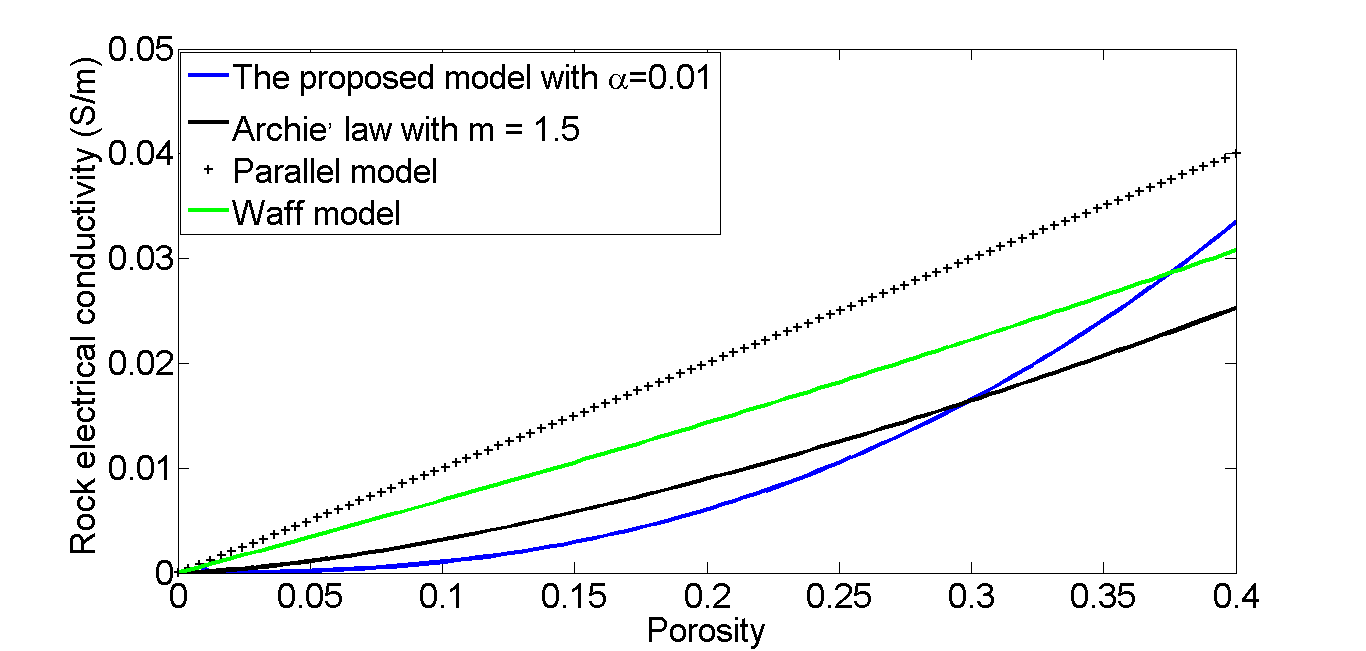}
  \caption{The variation of the electrical conductivity of porous samples with porosity predicted from different models with $\sigma_w$ = 0.1 S/m. For the proposed model shown by Eq. (\ref{equa30}), the grain diameter is taken as 50$\times10^{-6}$ m and $\alpha$ is taken as 0.01.}\label{Figure3}
\end{figure}

The proposed model indicated by Eq. (\ref{equa30}) for the case of negligible surface conductivity is applied to predict the variation of the electrical conductivity of porous samples with porosity and compared with other models such as the Archie model \cite[]{Archie1942}, parallel model \cite[]{Gueguen1994}, Waff model \cite[]{doi:10.1029/JB079i026p04003}. Table \ref{table2} lists some of the models for the electrical conductivity of porous media used in this work for comparison including the models given by \cite{Pride,doi:10.1029/98JB02125}. The prediction from the models are performed with $\sigma_w$ = 0.1 S/m. The proposed model is applied with the grain diameter of 50$\times10^{-6}$ m and $\alpha$ = 0.01. The comparison between those models is shown Fig. \ref{Figure3}. It is seen that all the models have the similar behavior and indicate the increase of electrical conductivity of porous media with increasing porosity. 

\begin{table*}
\caption{The parameters used in the proposed model to compare experimental data from different sources. Symbols of $d$ ($\mu$m), $\phi$ (no units), $\alpha$ (no units), $\sigma_w$ (S/m) and $\Sigma_s$ (S) stand for the grain diameter, porosity, ratio of minimum and maximum radius, fluid electrical conductivity and surface conductance of the fluid saturated porous samples, respectively}
\begin{center}
\begin{tabular}{c c c c c c c}
  \hline
  Sample&$d$ ($\mu$m) &$\phi$ (no units) & $\alpha$ (no units)& $\sigma_w$ (S/m)& $\Sigma_s$ (S)& source\\
  \hline
  S1a& 56  & 0.4 & 0.01& $10^{-4}$ to 0.1 & 0.5$\times10^{-9}$&\cite{Boleve} \\
	S2& 93  & 0.4 & 0.01& $10^{-4}$ to 0.1 & 0.5$\times10^{-9}$&\cite{Boleve} \\
  S3& 181  & 0.4 & 0.01& $10^{-4}$ to 0.1 & 0.5$\times10^{-9}$&\cite{Boleve} \\
  S4& 256  & 0.4 & 0.01& $10^{-4}$ to 0.1 & 0.5$\times10^{-9}$&\cite{Boleve} \\
  S5& 512  & 0.4 & 0.01& $10^{-4}$ to 0.1 & 0.5$\times10^{-9}$&\cite{Boleve} \\
  S6& 3000  & 0.4 & 0.01& $10^{-4}$ to 0.1 & 0.5$\times10^{-9}$&\cite{Boleve} \\
	\hline
	SW& 106  & 0.34 & 0.01& $10^{-5}$ to 1 & 1.0$\times10^{-9}$&\cite{Wildenschild2000} \\
	\hline
	FriedS& 50, 100  & 0.35-0.45 & 0.001& unused & 0 &\cite{doi:10.1029/2001WR000746} \\
	&  200, 500  &  & &  &  & \\
	\hline
	ReS& 250 & 0.04-0.23 & 0.0001& unused & unused &\cite{Revilgeo2014-0036}\\
	\hline
\end{tabular}
\end{center}
\label{table1}
\end{table*}

\subsection{Effect of the pore water electrical conductivity}
\label{ECeffect}

To compare the proposed model with experimental data available in literature, the input parameters corresponding to each sample are listed in Table \ref{table1}


\begin{figure}
\centering
  \begin{tabular}{@{}cc@{}}
    \includegraphics[width=.50\textwidth]{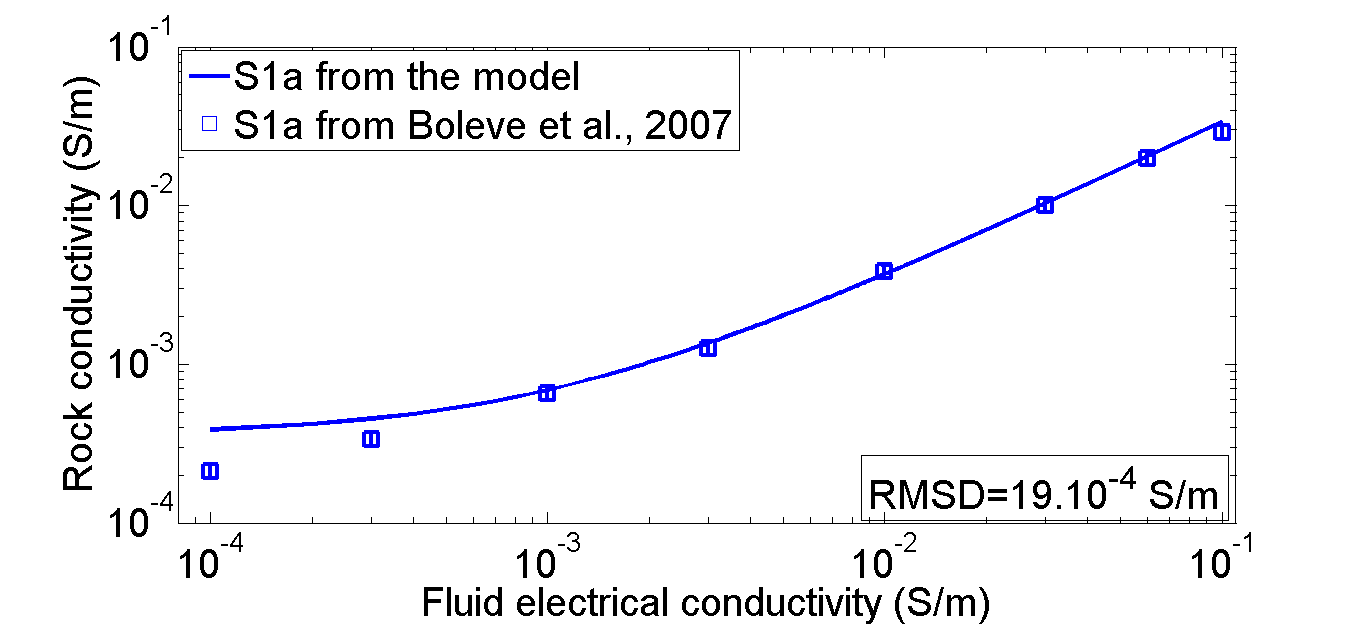} &
    \includegraphics[width=.50\textwidth]{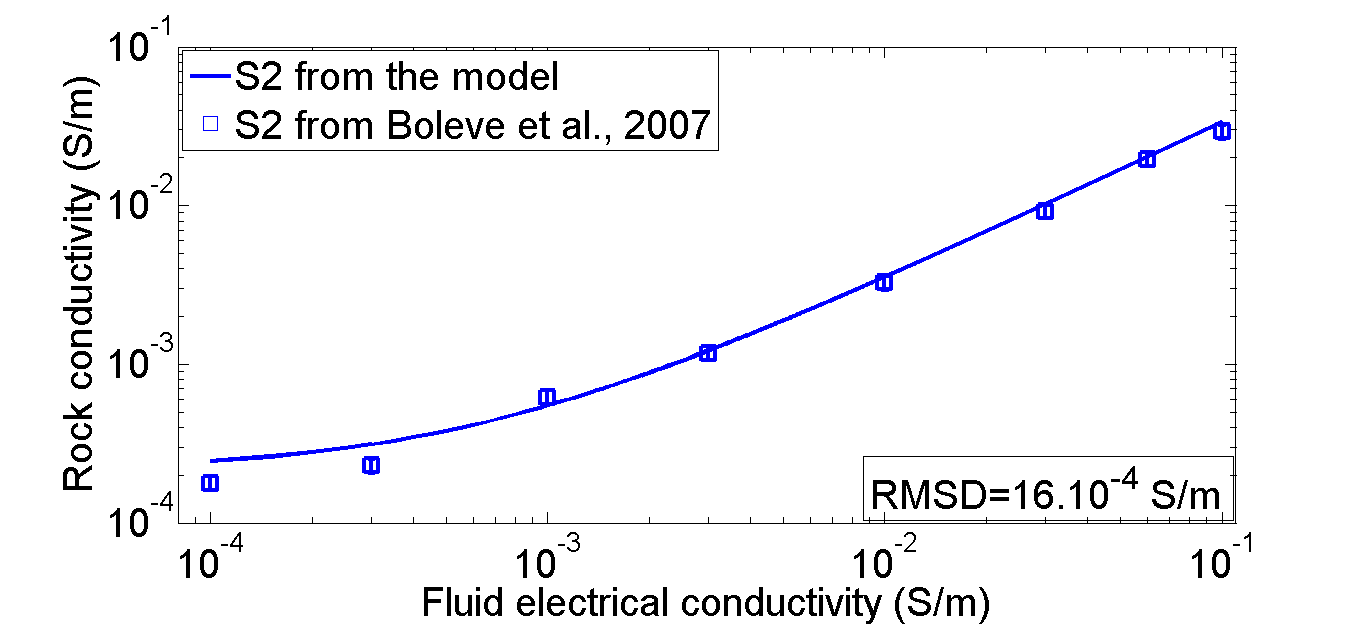}   \\
    \includegraphics[width=.50\textwidth]{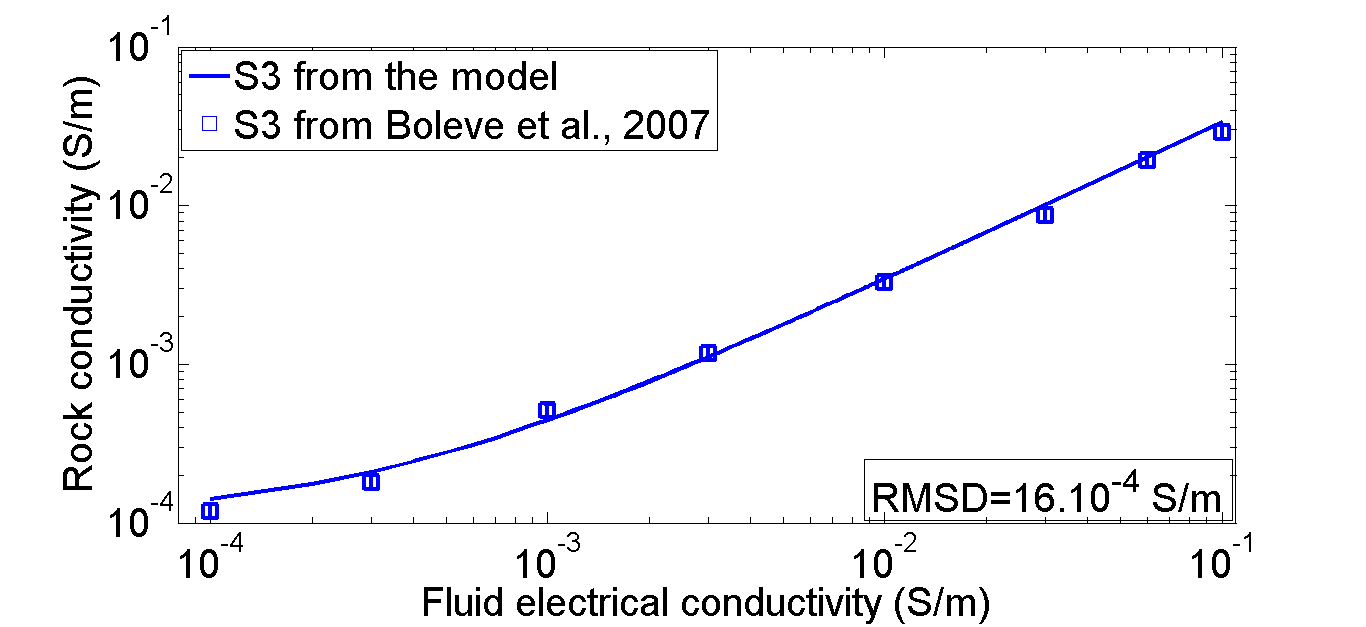} &
    \includegraphics[width=.50\textwidth]{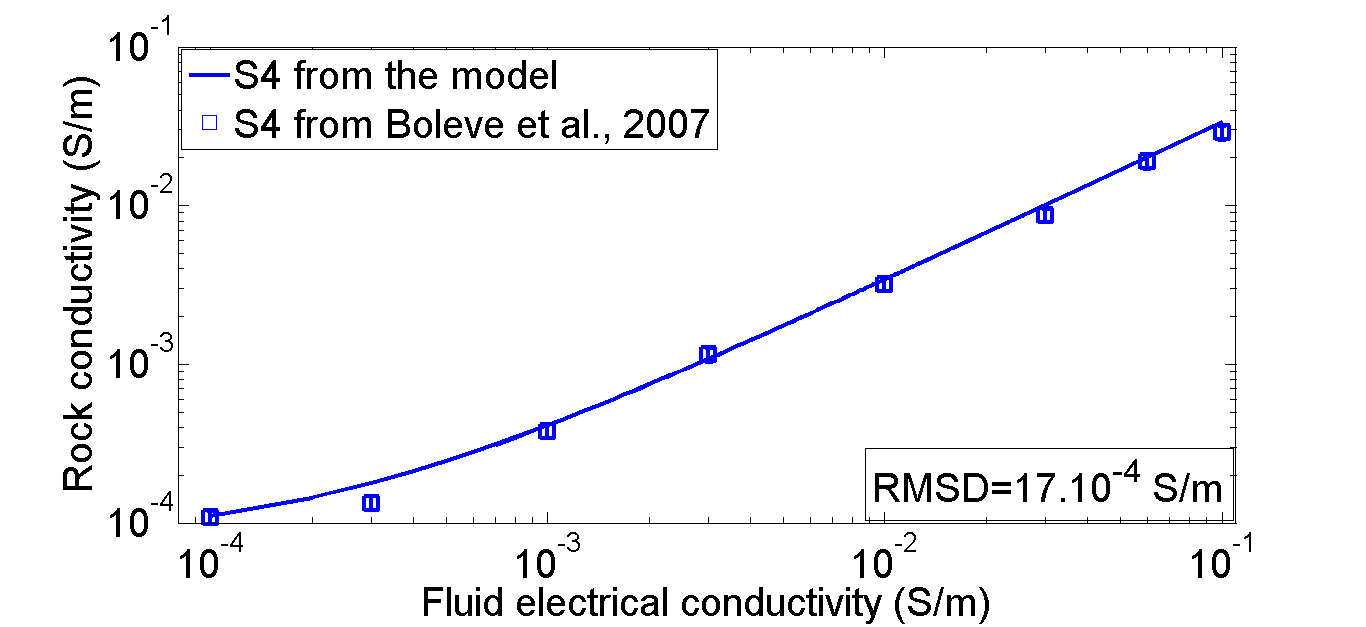} \\
		\includegraphics[width=.50\textwidth]{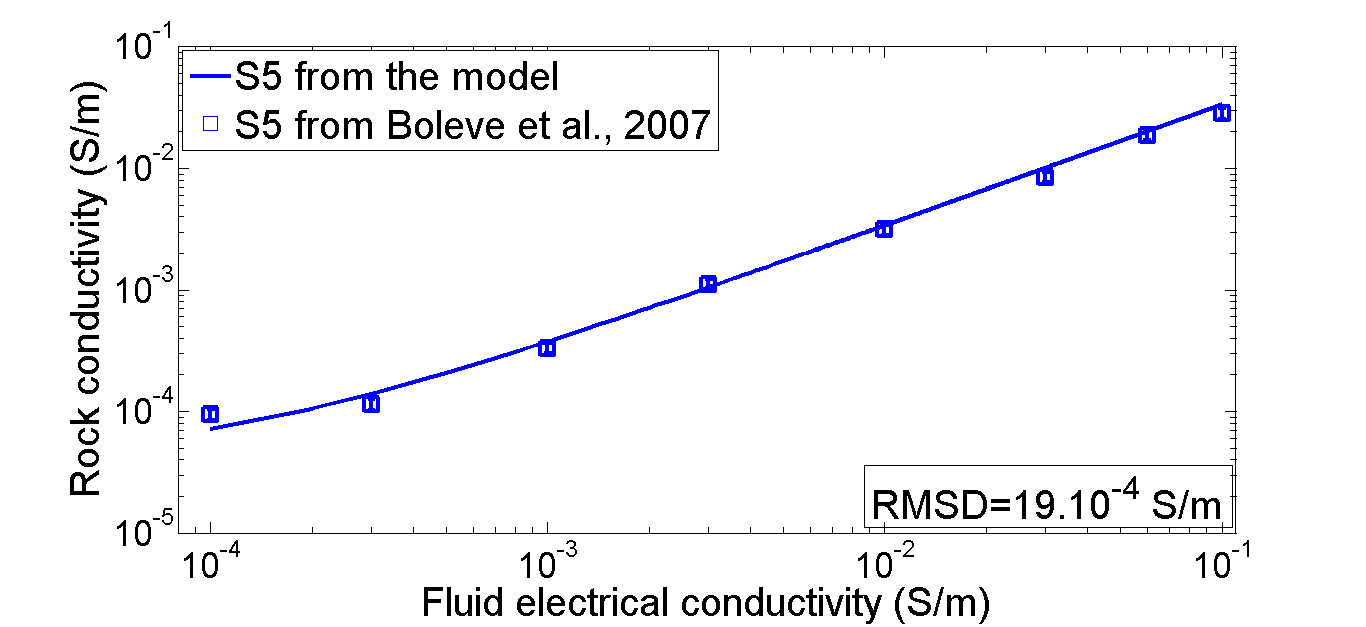} &
    \includegraphics[width=.50\textwidth]{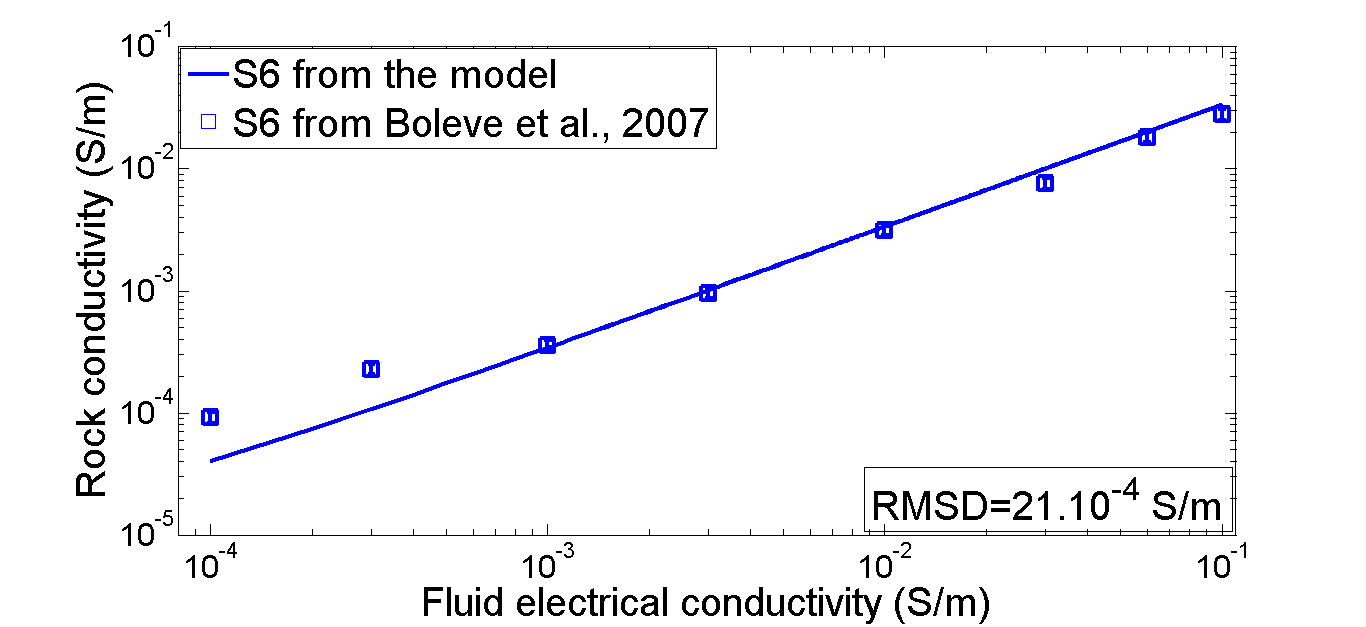} \\
  \end{tabular}
  \caption{Electrical conductivity of different packs of glass bead versus the electrical conductivity of the fluid. The symbols are obtained from the work of \cite{Boleve}. The solid lines are from the proposed model presented by Eq. (\ref{equa22}) with parameters given in Table \ref{table1}.}\label{Figure4}
\end{figure}

Fig. \ref{Figure4} shows the dependence of the electrical conductivity of saturated porous rocks as a function of the pore fluid electrical conductivity for six glass bead packs of different grain diameters experimentally obtained from \cite{Boleve} (see the symbols) and the prediction from the model presented by Eq. (\ref{equa22}) (the solid lines). Mean grain size  of six glass bead packs denoted by S1a, S2, S3, S4, S5 and S6 are 56 $\mu$m, 93 $\mu$m, 181 $\mu$m, 256 $\mu$m, 512 $\mu$m and 3000 $\mu$m, respectively (see Table \ref{table1}). The measured porosity of the packs was reported to be $\phi$ = 0.40 irrespective of the size of the glass beads \cite[]{Boleve}. By fitting experimental data shown in Fig. \ref{Figure4}, the surface conductance is found to be $\Sigma_s$ = 0.5$\times10^{-9}$ S for all samples. This value is of the same order of magnitude as that reported in literature for the silica surface in contact with the NaCl electrolyte (e.g., $\Sigma_s$ = 8.9$\times10^{-9}$ S \cite[]{Revil1998} or $\Sigma_s$ = 4.0$\times10^{-9}$ S \cite[]{Glover2010} or $\Sigma_s$ = 5.3$\times10^{-9}$ S \cite[]{Bull1932} or $\Sigma_s$ = 1.4$\times10^{-9}$ S \cite[]{Lorne1999}). The root-mean-square deviation (RMSD) calculated for all samples S1a, S2, S3, S4, S5 and S6 are 19$\times10^{-4}$ S/m, 16$\times10^{-4}$ S/m, 16$\times10^{-4}$ S/m, 17$\times10^{-4}$ S/m, 19$\times10^{-4}$ S/m and 21$\times10^{-4}$ S/m, respectively. The results show that the predictions from the model are in very good agreement with the experimental data. The model is able to reproduce the main trend of experimental data and especially at high fluid electrolyte concentration and large grain size. The misfit of the model with experimental data at low electrolyte concentration can be explained by the possible exchange of the electrical current between the electrical double layer and bulk fluid \cite[]{Daigle2015JB}. As seen in Fig. \ref{Figure4}, at high fluid electrical conductivity there is a linear dependence. The reason is that at high fluid electrical conductivity or large grain size, the electrical conductivity of saturated porous samples $\sigma$ is linearly related to the fluid electrical conductivity $\sigma_w$ as presented in Eq. (\ref{eq1}) or Eq. (\ref{equa30}), that is $Du << 1$.

\begin{figure}
  \centering
  \includegraphics[width=12.0cm]{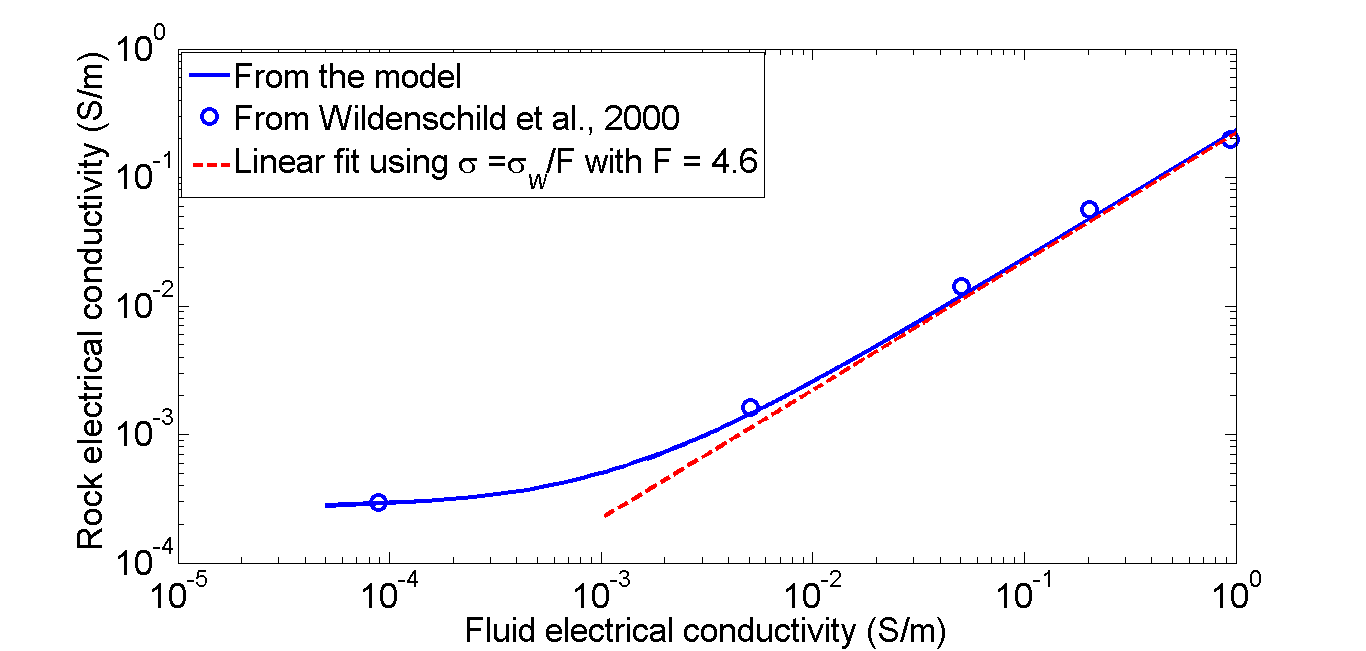}
  \caption{Electrical conductivity of a porous sample versus the electrical conductivity of the fluid. The symbols are obtained from \cite{Wildenschild2000}. The solid line is predicted from the model indicated by Eq. (\ref{equa22}) with parameters given in Table \ref{table1}. The RMSD is 90$\times10^{-4}$ S/m. The dashed line is linear fitting using the relation $\sigma$=$\sigma_w$/$F$ with $F$ = 4.6.}\label{Figure5}
\end{figure}

The dependence of the electrical conductivity of another saturated sand pack (denoted by SW) as a function of the pore fluid electrical conductivity from a different source \cite[]{Wildenschild2000} is also shown in Fig. \ref{Figure5}. The symbols are from the reported data and the solid line is predicted from the model with the parameters given in Table \ref{table1} in which the mean diameter of grains of a sand pack is deduced from \cite{Glover2009}

\begin{equation}
\label{equa32}
\begin{split}
d=2\theta\sqrt{8kF}\\
\end{split}
\end{equation}
where $\theta$ is the theta transform (unitless) and equal to 3.436 for spherical grain samples, $k$ and $F$ are the permeability and formation factor of porous samples, respectively ($k$ = 6.16$\times10^{-12}$ m$^2$ and $F$ = 4.9 for the sample SW as reported in \cite{Wildenschild2000}). Consequently, $d$ is calculated to be 106 $\mu$m. The linear trend of the curve (the dashed line) can be used to determine the formation factor. One can obtain $\sigma$=$\sigma_w$/4.6 for the dashed line in Fig. \ref{Figure5}. Consequently, the formation factor is 4.6. This theoretically predicted value is in good agreement with the measured value of 4.9 from \cite{Wildenschild2000}. The RMSD in this case is 90$\times10^{-4}$ S/m that is larger than those of consolidated samples as reported above.

\begin{figure}
\centering
  \begin{tabular}{@{}cc@{}}
    \includegraphics[width=.52\textwidth]{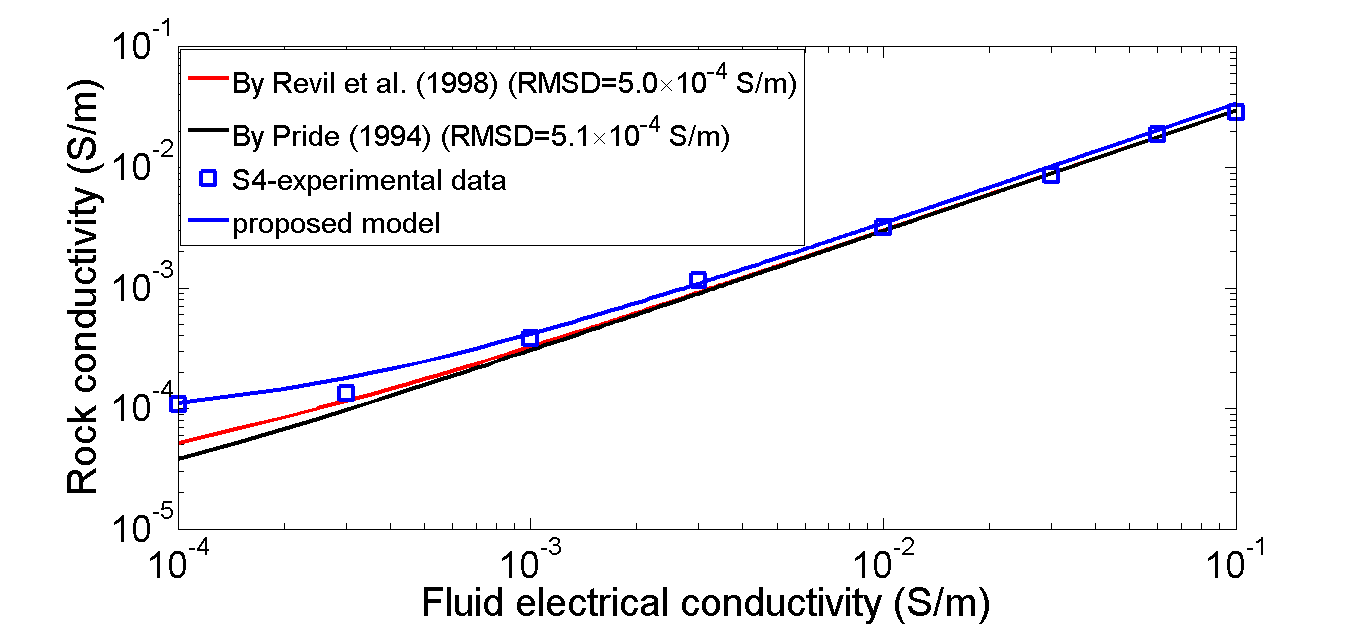} &
    \includegraphics[width=.52\textwidth]{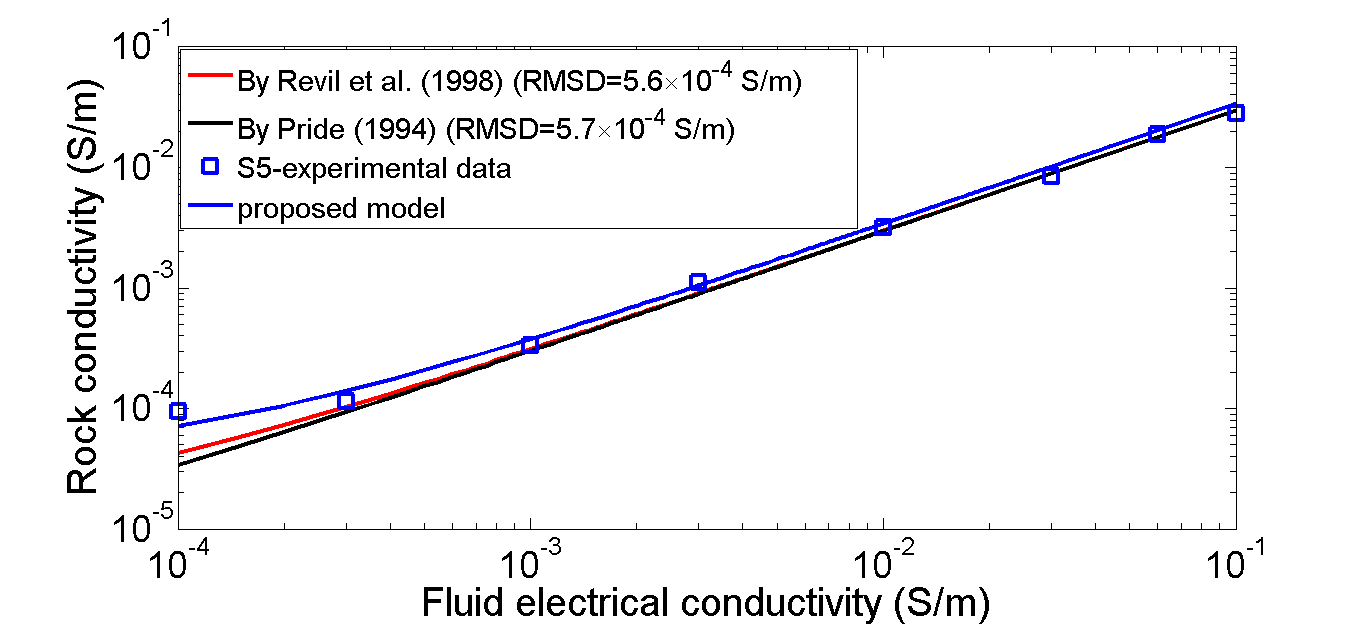} \\
  \end{tabular}
  \caption{The variation of electrical conductivity of saturated porous media with the electrical conductivity of the fluid predicted from the proposed model indicated by Eq. (\ref{equa22}) and other models given by \cite{Pride,doi:10.1029/98JB02125}.}\label{Figure6}
\end{figure}

The variation of electrical conductivity of saturated porous media with the electrical conductivity of the fluid predicted from the model is also compared with other models available in literature \citep[e.g.,][]{Pride,doi:10.1029/98JB02125} (see Table \ref{table2}) as shown in Fig. \ref{Figure6}. Input parameters for modeling are the same as those used in Fig. \ref{Figure4} for the representative samples S4 and S5 whose data are also shown in Fig. \ref{Figure6}. Additionally, $t^f_{(+)}$ is taken as 0.38 corresponding to NaCl solution \cite[]{doi:10.1029/98JB02125}, $\sigma_s$ is calculated by 6$\Sigma_s$/$d$ ($d$ is grain diameter) \cite[]{doi:10.1029/2006WR005131,Boleve} and $F$ is taken as 3.4 for both samples S4 and S5 \cite[]{Boleve}. The RMSD of the fits from \cite{doi:10.1029/98JB02125} and \cite{Pride} for samples S4 and S5 are 5.0$\times10^{-4}$ S/m and 5.1$\times10^{-4}$ S/m; 5.6$\times10^{-4}$ S/m and 5.7$\times10^{-4}$ S/m, respectively. It is shown that the proposed model is in good agreement with those given by \cite{Pride} and \cite{doi:10.1029/98JB02125}, especially at high fluid electrical conductivity (Dukhin number $Du$ $\leq$ 0.004). 

\subsection{Effect of the textural parameters of the porous medium}
\label{texturaleffect}

\begin{figure}
  \centering
  \includegraphics[width=12.0cm]{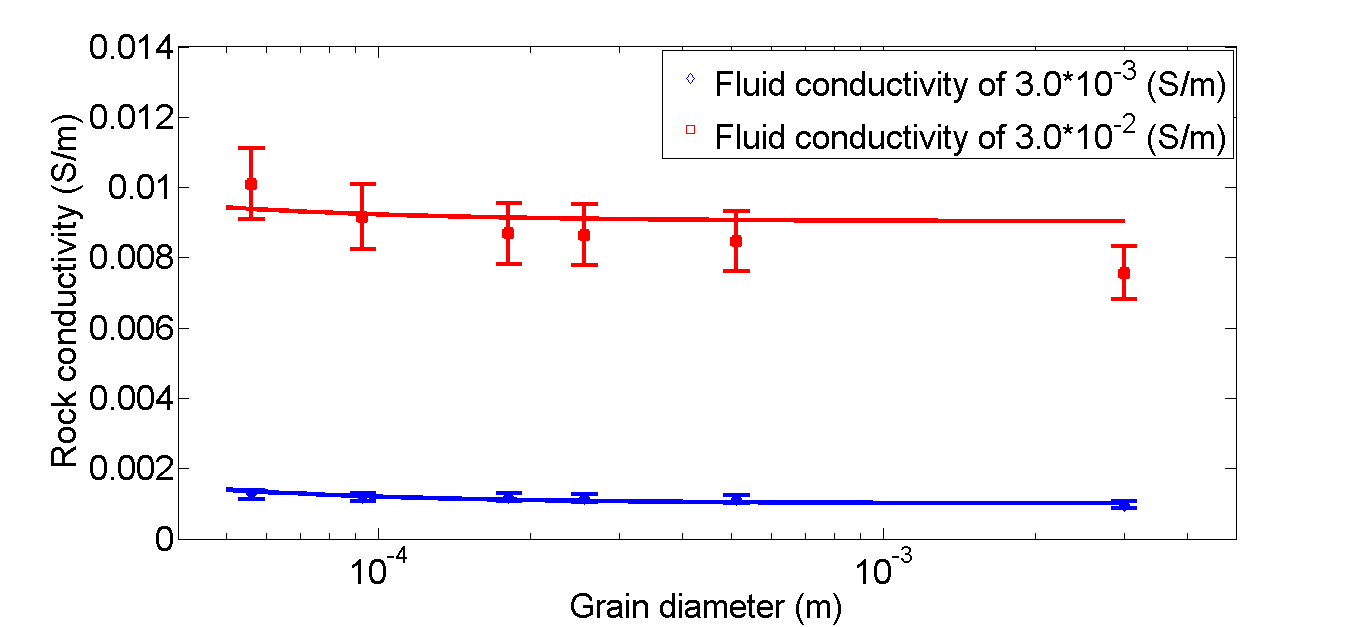}
  \caption{Electrical conductivity of porous samples versus the grain diameter of the samples at two different electrical conductivity of the fluid (3.0$\times10^{-3}$ and 3.0$\times10^{-2}$ S/m). Mean grain size of six glass bead packs are 56, 93, 181, 256, 512, and 3000 $\mu$m, respectively. The porosity of all of the packs is 0.4. The symbols are obtained from \cite{Boleve}. The solid lines are predicted from the model indicated by Eq. (\ref{equa22}) with parameters given in Table \ref{table1}.}\label{Figure7}
\end{figure}

The dependence of the electrical conductivity of saturated porous media on the grain diameter is also predicted using the procedure mentioned in section 4.2. The prediction is shown by the solid lines in Fig. \ref{Figure7} for two different fluid conductivities (3.0$\times10^{-3}$ S/m and 3.0$\times10^{-2}$ S/m). The theoretical results are then compared with the measured values obtained from \cite{Boleve} (see symbols). It should be noted that the input parameters for modeling are reported in Table \ref{table1} for \cite{Boleve}. The uncertainties of the measured electrical conductivities of porous samples and grain diameter are reported to be roughly 10 $\%$ by \cite{Boleve}. Therefore, a $\pm$ 10 $\%$ error is used for the experimental data shown in Fig. \ref{Figure7}. The comparison shows that the model is able to reproduce the main trend of the experimental data. However, it is seen that the fit at lower fluid conductivity is not as good as at higher one. The reason is that the assumption of the conduction current path in parallel may be not truly valid at low electrolyte concentration \cite[]{Daigle2015JB}.

\begin{figure}
  \centering
  \includegraphics[width=12.0cm]{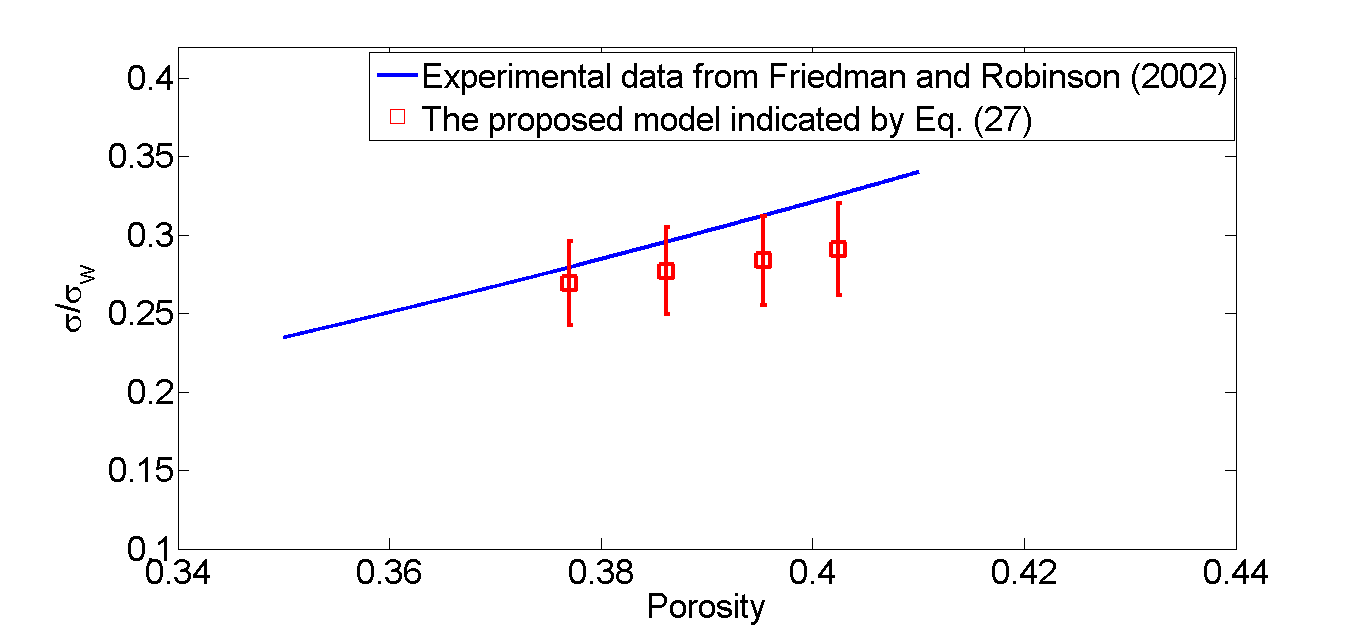}
  \caption{Ratio of the electrical conductivity of the glass bead sample to that of the electrolyte as a function of the porosity. Mean grain size of four samples are 50, 100, 200, and 500 $\mu$m, respectively. The symbols are obtained from \cite{doi:10.1029/2001WR000746} with $\pm$ 10 $\%$ uncertainty. The solid line is from the proposed model given by Eq. (\ref{equa30}) with parameters given in Table \ref{table1}.}\label{Figure8}
\end{figure}

Figure \ref{Figure8} shows the ratio of the electrical conductivity of the glass bead samples to that of the electrolyte as a function of the porosity experimentally obtained from \cite{doi:10.1029/2001WR000746} (see symbols) for four samples of spherical glass beads with diameters of 50, 100, 200 and 500 $\mu$m denoted by FriedS in Table \ref{table1}. This measured result can be explained by the model (Eq. (\ref{equa30})) with input parameters given in Table \ref{table1}. The surface conductivity is negligible as stated in \cite{doi:10.1029/2001WR000746}. A $\pm$ 10 $\%$ error is used for the experimental data shown in Figure \ref{Figure8}. The result shows that the prediction of the model is close to the experimental data.

\subsection{Formation factor and hydraulic tortuosity}
\label{Formationfactor}

\begin{figure}
  \centering
  \includegraphics[width=12.0cm]{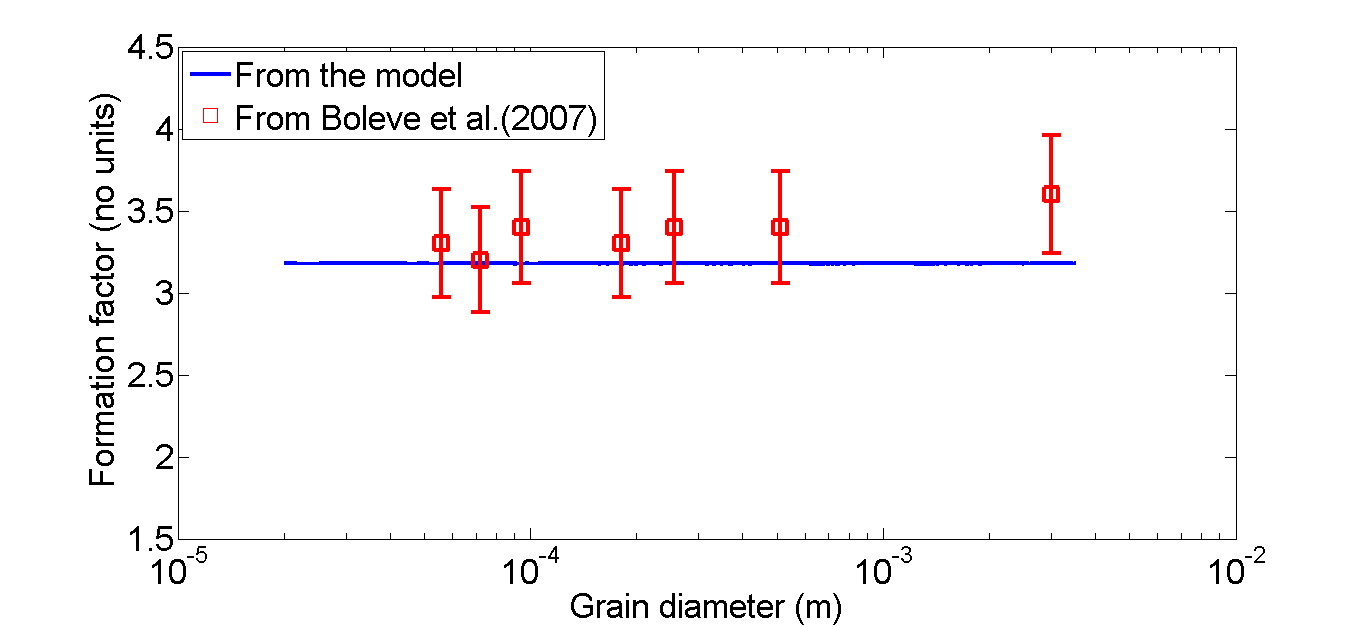}
  \caption{The variation of the formation factor with the grain diameter. The symbols are obtained from \cite{Boleve}. The solid line is predicted from the model indicated by Eq. (\ref{equa31}) with parameters given in Table \ref{table1} for the source of \cite{Boleve} with $\pm$ 10 $\%$ uncertainty on the experimental data.}\label{Figure9}
\end{figure}

The variation of the formation factor of the porous media with the grain diameter can be predicted from Eq. (\ref{equa31}) as shown in Fig. \ref{Figure9} (the solid line). The parameters for modeling are $\alpha$ = 0.01 and $\phi$ = 0.40 that are compatible with glass bead packs reported in \cite{Boleve} (see Table \ref{table1}). The formation factor measured by \cite{Boleve} for different grain diameters is also shown by the symbols in Fig. \ref{Figure9}. It is seen that the model is in good agreement with experimental data. According to Archie model \cite[]{Archie1942}, the formation factor $F$ is linked to the porosity $\phi$ ($\phi$ = 0.40 irrespective of the size of the glass beads as stated in \cite{Boleve}) by $F = \phi^{-m}$. For unconsolidated samples made of perfect spheres, the exponent should be constant (1.5) \cite[]{Sen}. Consequently, $F$ is predicted to be independent of the grain size. This constant value of formation factor predicted by the model in Eq. (\ref{equa31}) ($F$ = 3.21) is in very good agreement with the value measured by \cite{Boleve} (3.4$\pm$0.2) for glass beads.

\begin{figure}
  \centering
  \includegraphics[width=12.0cm]{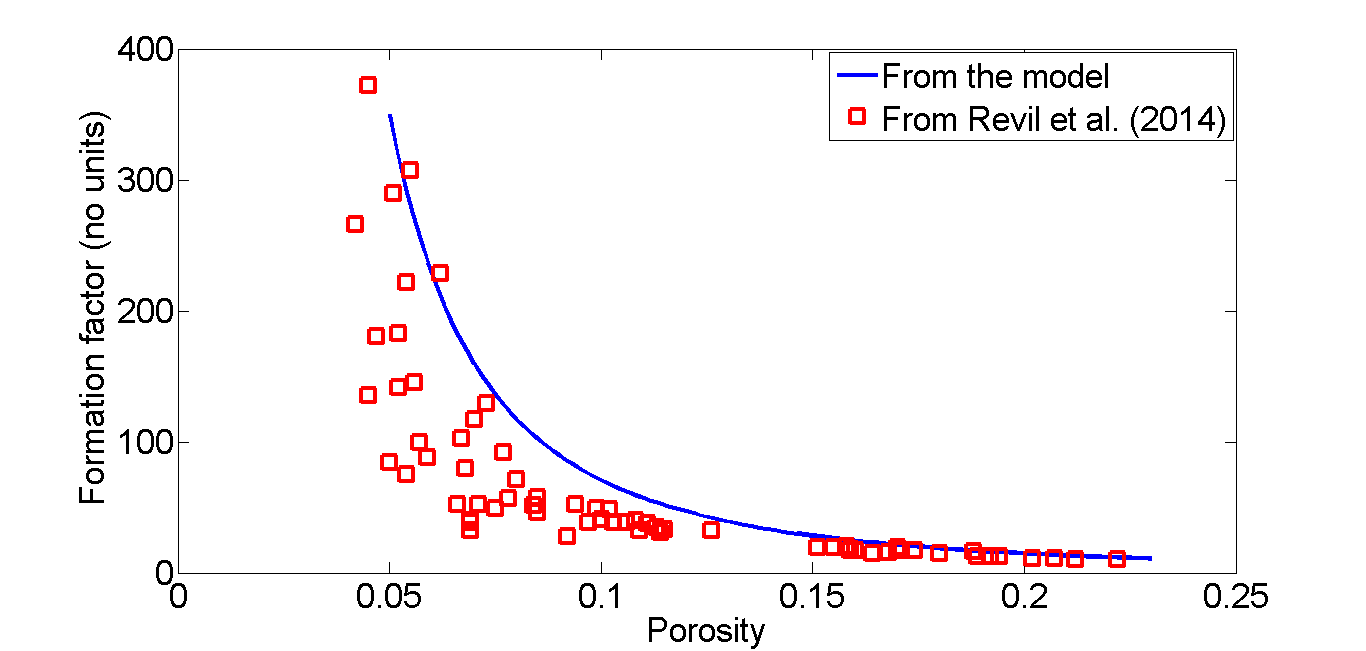}
  \caption{The variation of the formation factor with with porosity. The symbols are obtained from \cite{Revilgeo2014-0036}. The solid line is predicted from the model indicated by Eq. (\ref{equa31}) with input parameters given in Table \ref{table1}.}\label{Figure10}
\end{figure}

The variation of the formation factor of the porous media with porosity experimentally obtained for a set of core samples of the Fontainebleau sandstone \cite[]{Revilgeo2014-0036} is shown in Fig. \ref{Figure10} (see the symbols). The main experimental trend can be explained by the model indicated by Eq. (\ref{equa31}) (the solid line). The value of $\alpha$ = 0.0001 is used for modeling because of the best fit and that value is also used to fit experimental data for Fontainebleau sandstone \citep[e.g.,][]{Liang2014}. The mean grain diameter is $d$ = 250 $\mu$m obtained from CT scans in \cite{Revilgeo2014-0036} (see sample ReS in Table \ref{table1}). It is indicated that the model can produce the main behavior of the experimental data given by \cite{Revilgeo2014-0036}. However, a deviation is observed at very low porosity (less than 15 $\%$). The reason may be that the model is developed based on porous media made up of mono-sized spherical grains \cite[]{Liang2015}. However, for the consolidated samples of the Fontainebleau sandstone, the rock texture consists of mineral grains of various shapes and sizes and its pore structure is extremely complex. The formula for the fractal dimension $D_f$, the maximum radius $r_{max}$ determined from the grain diameter, porosity may not be suitable. Additionally, another reason for a deviation is the variation of $\alpha$ from sample to sample (for modeling, $\alpha$ = 0.0001 is applied for all samples).

Based on Eq. (\ref{equa31}), the effective electrical tortuosity is defined as \citep[e.g.,][]{Clennell299}
\begin{equation}
\label{equa33}
(\tau^{eff}_e)^2=F\phi=(\tau^{eff}_g)^2\frac{(D_{\tau}-D_f+1)(1-\alpha^{3-D_{\tau}-D_f})}{(3-D_{\tau}-D_f)(1-\alpha^{D_{\tau}-D_f+1})}\\
\end{equation}
Eq. (\ref{equa33}) is used to predict the effective electrical tortuosity of 12 samples of glass beads with different diameters reported in \cite{Glover2010}. The input parameters for modeling and predicted values for all samples are listed in Table \ref{table3}. It is seen that the mean value over 12 samples is around 1.11 and that is smaller than the hydraulic tortuosity of those samples predicted by \cite{doi:10.1002/2017JB014873} using different approaches ($\tau_h^{av}$=(1.2+1.4)/2=1.3). The result is in good agreement with \citep{GhanbarianBehzad2013}. This work provides a possible link between different properties. For example, the hydraulic conductivities of porous media can be deduced from  electrical conductivities \citep[e.g.,][]{doi:10.1029/2008WR007309,doi:10.1029/2009WR008552,NIU2015475}. 

\begin{table*}
\caption{Input parameters for modeling and predicted hydraulic tortuosities. Grain diameter ($d$ in $\mu$m) and porosity ($\phi$) are taken from \cite{Glover2010}}
\begin{center}
\begin{tabular}{c c c c c }
  \hline
 Sample number &$d$  & $\phi$ & $\alpha$ & $\tau^{eff}_e$\\
  \hline
 1 & 1.05& 0.411 & 0.01 & 1.072\\
	2 &2.11 &  0.398 & 0.01 & 1.098\\
	3 & 5.01& 0.385 &  0.01& 1.125\\
	4 &  11.2& 0.401  & 0.01 & 1.092\\
	5 &21.5 & 0.383 & 0.01 & 1.128\\
	6 & 31.0& 0.392  & 0.01 & 1.110\\
	7 & 47.5& 0.403 &  0.01& 1.088\\
	8 &104  &  0.394 &  0.01& 1.106\\
	9 &181 & 0.396 &  0.01& 1.102\\
	10 & 252&  0.414 &  0.01& 1.067\\
	11 & 494&  0.379&  0.01& 1.137\\
	12 &990 &  0.391 &  0.01& 1.111\\
	mean & &   &  & 1.110\\
	\hline
\end{tabular}
\end{center}
\label{table3}
\end{table*}

\clearpage
       
\section{Conclusions}
\label{Conclusions}

A physically based model for estimating electrical conductivity of saturated porous media has been developed in this work. The model is derived assuming that the porous media can be represented by a bundle of tortuous capillary tubes with a fractal pore-size distribution. The proposed model is explicitly related to electrical conductivity of the pore liquid and the microstructural parameters of a porous medium ($D_f$, $D_{\tau}$, $\phi$, $\alpha$, $r_{max}$). Therefore, the model can reveal more mechanisms affecting the electrical conductivity of porous media than other models available in literature. From the model, the expressions for the formation factor and hydraulic tortuosity are also obtained.

The model's sensitivity to its parameter is first tested. Then, it is compared with previously published models and experimental data. The proposed model is consistent with previous ones from the literature. The results predicted by the model are in very good agreement with experimental data, especially for unconsolidated samples. The proposed model is the first fractal theory based one to take the surface conductivity into account. This simple analytical model opens-up new possibilities for predicting the electrical conductivity of porous media and link it with hydraulic conductivity. The analytical development of a model for partially saturated porous media using the presented approach will be carried out in the near future.












\end{document}